\documentclass[aps,showpacs,nofootinbib,superscriptaddress,preprintnumbers]{revtex4}

\usepackage{graphicx}
\usepackage{amsmath}
\usepackage{bm}
\usepackage{color}
\newcommand{\de}{d}
\newcommand{\tr}{\mbox{Tr}\,}
\newcommand{\ii}{i}
\newcommand{\Pslash}{\rlap{\,/} P}
\newcommand{\kslash}{\rlap{/} k}
\newcommand{\lslash}{\rlap{/} l}

\begin{document}

\title{
Collins fragmentation function for pions and kaons in a spectator model
}

\author{Alessandro Bacchetta}
\email{alessandro.bacchetta@desy.de}
\affiliation{Theory Group, Deutsches Elektronen-Synchroton DESY, \\
D-22603 Hamburg, Germany}

\author{Leonard P. Gamberg}
\email{lpg10@psu.edu}
\affiliation{Department of Physics, Penn State University, Berks,\\
 Reading, PA 19610, USA}

\author{Gary R. Goldstein}
\email{gary.goldstein@tufts.edu}
\affiliation{Department of Physics and Astronomy, Tufts University, \\
 Medford, MA 02155, USA}

\author{Asmita Mukherjee}
\email{asmita@phy.iitb.ac.in}
\affiliation{Physics Department, Indian Institute of Technology Bombay, \\
Powai, Mumbai 400076, India}

\begin{abstract}
We calculate the Collins fragmentation function in the framework of a
spectator
model with pseudoscalar pion-quark coupling and a Gaussian form
factor at the vertex. 
We determine the model parameters by fitting the
unpolarized fragmentation function for pions and kaons. 
We show that the Collins function for the pions in
this model is in reasonable agreement with 
recent parametrizations obtained by
fits of the available data. In addition, we compute for the first time 
the Collins function for the kaons.
\end{abstract}




\preprint{DESY 07-105}

\maketitle

\section{Introduction}
\label{s:intro}

The Collins fragmentation function~\cite{Collins:1993kk} 
 measures how the orientation of the
quark spin influences the direction of emission of hadrons in the
fragmentation process and can thus be used as a quark spin analyzer.
It contributes to 
several single spin
asymmetries (SSA) in hard processes, such as semi inclusive deep inelastic
scattering (SIDIS), 
$pp$ collisions and
$e^+e^-$ annihilation into hadrons. We shall henceforth use the term ``Collins
asymmetries'' to denote any asymmetry where the Collins function plays a role.

The first experimental evidence of a nonzero Collins function for pions 
came from
the measurement of a Collins asymmetry in SIDIS on a proton target by the
HERMES collaboration~\cite{Airapetian:2004tw}.  The same asymmetry, but on a
deuteron target, was found to be consistent with zero by the COMPASS
collaboration~\cite{Ageev:2006da}. At the moment, the most
convincing evidence of a nonzero pion Collins function comes from the
measurements of a Collins asymmetry in $e^+e^-$
annihilation~\cite{Abe:2005zx}. First extractions of the pion Collins 
function were performed in Ref.~\cite{Vogelsang:2005cs,Efremov:2006qm}. 
A recent fit to SIDIS
and $e^+e^-$ annihilation allowed the simultaneous extraction of the Collins
fragmentation function and of the transversity parton distribution
function~\cite{Anselmino:2007fs}, 
clearly showing the importance of the Collins function as a tool to
investigate the structure of hadrons.
The kaon Collins function is at the moment unknown.

A few model calculations of the Collins function for pions 
have been presented in the
literature~\cite{Bacchetta:2001di,Bacchetta:2002tk,Gamberg:2003eg,Bacchetta:2003xn,Amrath:2005gv}
and used to make predictions and/or compare to available
data~\cite{Schweitzer:2003yr,Gamberg:2003eg,Gamberg:2003pz,Gamberg:2004wt}.
However, the
above calculations have been found to be inadequate to describe the
data.

The aim of the present work is to show that a Collins function in reasonable
agreement with the available parametrizations can be obtained in 
a model with pseudoscalar pion-quark 
coupling and Gaussian form
factors at the pion-quark vertex. We also present, for the first time, 
 the Collins function 
for the fragmentation of quarks into kaons. This calculation is
relevant for the interpretation of recent kaon measurements done at
HERMES~\cite{Diefenthaler:2006vn} as well as COMPASS~\cite{Bradamante:2007ex} and for
future measurements at BELLE and JLab.

\section{Model calculation of the unpolarized fragmentation function}

In the fragmentation process, 
 the probability to produce hadron $h$ from a transversely
polarized quark $q$, in, e.g., the $q \bar{q}$ rest frame if the fragmentation
 takes place in 
$e^+ e^-$ annihilation,
is given by (see, e.g., \cite{Bacchetta:2004jz})
\begin{equation} 
D_{h/q^\uparrow}(z, K_T^2) =  D_1^q(z, K_T^2) + H_1^{\perp q}(z, K_T^2) \,
    \frac{(\hat{\bm k} \times {\bm K}_{T}) \cdot {\bm s}_q}{z M_h},
\end{equation}
where $M_h$ the hadron mass, $k$ is the momentum of the quark, $s_q$ its
spin vector,  $z$ is the light-cone momentum
fraction of the hadron with respect to the fragmenting quark, and
$K_{T}$ the component of the hadron's momentum transverse to $k$. 
$D_1^q$ is
the unintegrated unpolarized fragmentation function, while 
$H_1^{\perp  q}$ is the Collins function.
Therefore, $H_1^{\perp q} > 0$ corresponds to a preference of the
hadron to move to the left if the quark is moving away from the
observer and the quark spin is pointing upwards.

In accordance with factorization,
fragmentation functions can be calculated from the correlation
function~\cite{Bacchetta:2006tn}  
\begin{equation}
\begin{split} 
\Delta(z,k_T)  &= \frac{1}{2z} \int \de k^+ \Delta(k,P_h)
\\&
=\frac{1}{2z}\sum_X \, \int
  \frac{\de\xi^+  \de^2\bm{\xi}_T}{(2\pi)^{3}}\; e^{\ii k \cdot \xi}\,
    \langle 0|\, {\cal U}^{n_+}_{(+\infty,\xi)}
\,\psi(\xi)|h, X\rangle 
\langle h, X|
             \bar{\psi}(0)\,
{\cal U}^{n_+}_{(0,+\infty)}
|0\rangle \Bigr|_{\xi^-=0}\,,    
\label{e:delta}
\end{split} 
\end{equation} 
with $k^- = P_h^-/z$. A discussion on the structure of the Wilson lines, 
${\cal U}$, can be
found in Ref.~\cite{Bacchetta:2006tn}. Here, we limit ourselves to recalling
that in Refs.~\cite{Metz:2002iz,Collins:2004nx} it was shown that the 
fragmentation correlators are the same in both semi-inclusive DIS and $e^+e^-$
annihilation, as was also observed earlier in the context of a specific model
calculation~\cite{Metz:2002iz} similar to the one under consideration here.
In the rest of the article we shall utilize the Feynman gauge, in which
transverse gauge links at infinity give no contribution and can be
neglected~\cite{Ji:2002aa,Belitsky:2002sm,Boer:2003cm}. 

\begin{figure}
\includegraphics[width=5cm]{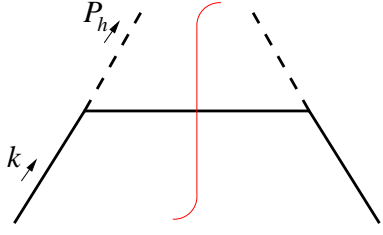}%
\caption{\label{treelevel} 
Tree-level diagram for quark to meson fragmentation
process.}    
\end{figure}

The tree-level diagram describing the fragmentation of a virtual (timelike) 
quark into a pion/kaon is shown in Fig.~\ref{treelevel}. 
In the model used here, the final
state $|h, X\rangle$ is described by the detected pion/kaon and
an on-shell spectator, with the quantum numbers of a quark and with mass
$m_s$.  
We take a pseudoscalar pion-quark coupling 
of the form $g_{q\pi} \gamma_5 \tau_i$, where $\tau_i$ are the generators of the
SU(3) flavor group. Our model is similar to the ones used in, e.g.,
Refs.~\cite{Ji:1993qx,Londergan:1996vf,Jakob:1997wg,Kitagawa:2000ji}. The most
important difference from previous calculations that included also the Collins
function, i.e., those in
Refs.~\cite{Bacchetta:2001di,Bacchetta:2002tk,Gamberg:2003eg,Bacchetta:2003xn,Amrath:2005gv},
is  that the mass of the spectator $m_s$ is not
constrained to be equal to the mass of the fragmenting quark. 

The fragmentation correlator at tree level, for the case $u\to \pi^+$, is
\begin{equation}
\begin{split} 
\Delta_{(0)}(k,p) & =
 - \frac{2\,g_{q\pi}^2}{(2\pi)^4}\,
 \frac{(\kslash + m)}{k^2 - m^2} \, \gamma_5  \, (\kslash - \Pslash_h +m_s)
\gamma_5 \,\frac{(\kslash + m)}{k^2 - m^2}\, 
 2\pi\,\delta\bigl((k-P_h)^2 -m_s^2\bigr) \, 
\end{split}
\end{equation} 
and, using the $\delta$-function to perform the $k^+$ integration,
\begin{equation}
 \Delta_{(0)}(z,k_T) = \frac{2\,g_{q\pi}^2}{32 \pi^3}\,
\frac{(\kslash + m)  \, (\kslash - \Pslash_h -m_s)  
 \,(\kslash + m)}{(1-z)P_h^-(k^2 - m^2)^2},
\label{D0}
\end{equation} 
where $k^2$ is
related to $k_T^2$ through the relation
\begin{equation} 
k^2=z k_T^2/(1-z)+ m_s^2/(1-z)+M_h^2/z
\label{e:k2kt2}
\end{equation} 
which follows from the on-mass-shell condition of the spectator quark of
mass $m_s$. We take $m$ to be the same for $u$
and $d$ quarks, but different for $s$ quarks. 
Isospin and charge-conjugation 
relations imply
\begin{gather} 
D_1^{u\to \pi^+}= D_1^{\bar{d}\to \pi^+} = D_1^{d\to \pi^-} = D_1^{\bar{u}\to
  \pi^-},
\\
D_1^{u\to K^+} = D_1^{\bar{u}\to K^-},
\\
D_1^{\bar{s}\to K^+} = D_1^{s\to K^-}.
\end{gather} 
For later purposes it is useful to spell out the relations coming
from isospin and charge-conjugation relations for the
unfavored functions
\begin{gather} 
D_1^{\bar{u}\to \pi^+} = D_1^{d\to \pi^+} = D_1^{\bar{d}\to \pi^-} = D_1^{u\to
  \pi^-} 
,
\label{unfisocharge1}
\\
D_1^{s\to \pi^+} = D_1^{\bar{s}\to \pi^+}= 
D_1^{s\to  \pi^-}= 
D_1^{\bar{s}\to \pi^-}
,
\label{unfisocharge2}
\\
D_1^{\bar{u}\to K^+} = D_1^{\bar{d}\to K^+}= D_1^{d\to K^+} = 
 D_1^{\bar{d}\to K^-}= D_1^{d\to K^-} = D_1^{u\to K^-}
,
\label{unfisocharge3}
\\
D_1^{s\to K^+} = D_1^{\bar{s}\to K^-} 
.
\label{unfisocharge4}
\end{gather} 
We assume the above relations hold for all fragmentation functions,
  in particular for the Collins function.

The unpolarized fragmentation function is projected from Eq.~(\ref{D0})
\begin{equation}
D_1(z, k_T^2)= \tr[\Delta_{0}(z, k_T) \gamma^+]/2
\end{equation} 
leading to the result
\begin{equation}
D_1(z, k_T^2)= \frac{g_{q\pi}^2}{8 \pi^3} 
\frac{\bigl[z^2 \bm{k}_T^2 + (z m + m_s-m)^2 \bigr]}
{z^3\,(\bm{k}_T^2 + L^2)^2},
\end{equation}
with
\begin{equation}
L^2= \frac{(1-z)}{z^2}\, M_h^2 +m^2 +\frac{m_s^2-m^2}{z}.
\end{equation}
In the limit $m_s=m$ and setting the form factor to $1$, our result for 
$D_1$ reduces to Eq.~(3) of Ref.~\cite{Amrath:2005gv} (multiplied by two because in that paper the results refer to $u \to \pi^0$).
The two nonzero kaon fragmentation functions $D_1^{u\to K^+}$ and
$D_1^{\bar{s}\to K^+}$ are given by the same functional
form, but with different masses $m$, $m_s$, $M_h$. 

The integrated unpolarized fragmentation function  is defined as
\begin{equation}
D_1(z)= \pi z^2  \int_0^\infty dk_T^2\, D_1 (z, k_T^2).
\label{integral}
\end{equation}
Here the integration is over the transverse momentum of the produced hadron
$K_T = -z k_T$ with respect to the quark direction, which is why
an extra factor of $z^2$ appears in the above equation.
The above integral is divergent. In Ref.~\cite{Amrath:2005gv}, a cutoff on
$k_T$ has been used. On the other hand, in Ref.~\cite{Gamberg:2003eg}, 
a Gaussian form
factor depending on $k_T^2$ has been introduced at the pion-quark
vertex, which effectively cuts off the higher $k_T$ region in the
integration. Similarly, 
in this work we use a Gaussian form factor of the form
\begin{align} 
g_{q\pi} &\mapsto \frac{g_{q\pi}}{z}\,
e^{-\frac{k^2}{\Lambda^2}}
&
\text{with}
&&
\Lambda^2 &= \lambda^2 z^\alpha (1-z)^\beta.
\label{e:ff}
\end{align}  
at the pion-quark vertex. Due to Eq.~\eqref{e:k2kt2}, the above form factor effectively cuts off the
higher $k_T$ region of the integration. The
form of the vertex is chosen merely on the basis of phenomenological
motivations, 
in order to reasonably reproduce the unpolarized fragmentation function. 
With the chosen form factor, the integration in Eq.~\eqref{integral} 
can be carried out analytically and yields:
\begin{equation}
\begin{split} 
D_1(z) =\frac{g_{q\pi}^2}{8 \pi^2}\,\frac{e^{-\frac{2 m^2}{\Lambda^2}}}{z^3
  L^2} \biggl[&(1-z)\Bigl((m_s-m)^2-M_h^2\Bigr)\,
                \exp\Bigl( -\frac{2 z L^2}{(1-z) \Lambda^2}\Bigr)
\\&+ 
 \biggl(z^2\Lambda^2-2z\Bigl((m_s-m)^2-M_h^2\Bigr)\biggr)\frac{L^2}{\Lambda^2}\,\Gamma\Bigl(0,\frac{2 z L^2}{(1-z) \Lambda^2}\Bigr)
\biggr],
\end{split} 
\end{equation} 
where the incomplete gamma function is, 
\begin{equation}
\Gamma(0,z) \equiv \int_z^\infty \frac{e^{-t}}{t}\, \de t.
\end{equation} 

The parameters of the model are 
$\lambda$, 
$\alpha$, $\beta$, 
together with the mass of the spectator $m_s$ and the mass of the
initial quark $m$. For the latter, we choose a constituent quark mass $m=0.3$
GeV for the $u$ and $d$ quarks, and $m=0.5$ GeV for the $s$ quark.
To fix the values of the other parameters, 
we performed a fit to the parametrization of
Ref.~\cite{Kretzer:2000yf} (NLO set) 
at the lowest possible scale, i.e., $Q_0=0.4$ GeV$^2$. The resulting values
for the parameters are
\begin{align} 
g_{q\pi}&= 4.78, 
&
\lambda&= 3.33 ~\mathrm{GeV},
&
\alpha&=0.5~\mathrm{(fixed)},
&
\beta&=0~\mathrm{(fixed)},
\end{align} 
which are common to both pion and kaon fragmentation functions. 
The only
parameters that change according to the type of fragmentation
function are
\begin{align}
u \to \pi^+&:
&
m_s&=0.792~\mathrm{GeV},
&
m=&0.3  ~\mathrm{GeV~(fixed)},
\\
u \to K^+&:
& 
m_s&=1.12 ~\mathrm{GeV},
&
m=&0.3  ~\mathrm{GeV~(fixed)},
\\
\bar{s} \to K^+&:
&
m_s&=0.559 ~\mathrm{GeV},
&
m=&0.5  ~\mathrm{GeV~(fixed)}.
\end{align} 
Obviously, also the mass of the hadron changes: we take $m_h=0.135$ GeV for
the pions and $m_h=0.494$ GeV for
the kaons. We remark that it is not possible to estimate the errors in the
parameters  in a meaningful way because the fragmentation functions
in Ref.~\cite{Kretzer:2000yf} have no error bands. It could be in principle
possible to use the recent parametrizations with error
bands~\cite{Hirai:2007cx},  
but the lowest scale they reach is 1 GeV$^2$, which we consider to be 
too high to compare to our model.

\begin{figure}
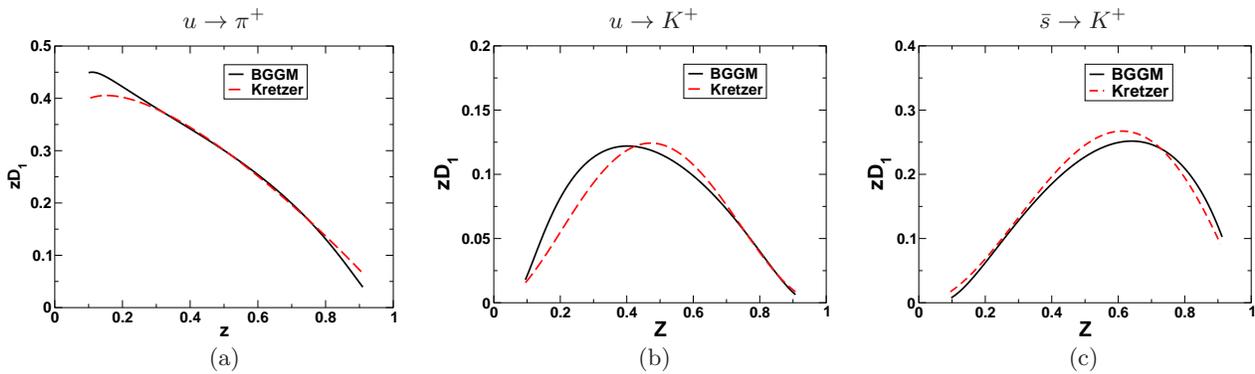

\begin{tabular}{ccccc}
$u \to \pi^+$& &$u \to K^+$& &$\bar{s} \to K^+$
\\
\hspace{-0.6cm}\includegraphics[height=4cm,clip]{zD1_upi+}
&
\hspace{0.8cm}
&
\hspace{-0.6cm}\includegraphics[height=4cm,clip]{zD1_uK+}
&
\hspace{0.8cm}
&
\hspace{-0.6cm}\includegraphics[height=4cm,clip]{zD1_sbK+}
\\
(a)& &(b)& &(c)
\end{tabular}
\caption{\label{D1} Unpolarized fragmentation function $zD_1(z)$ vs. $z$
for the fragmentation (a) $u \to \pi^+$, (b) $u \to K^+$, (c) 
$\bar{s} \to K^+$ in the spectator model (solid line), 
with parameters fixed from a fit to
the  parametrization of~\cite{Kretzer:2000yf} (dashed line).}    
\end{figure}
Fig.~\ref{D1} show the plots of the unpolarized fragmentation function
$D_1(z)$ multiplied by $z$ for $u \to \pi^+$, $u \to K^+$, and 
$\bar{s} \to K^+$. The parametrization
of \cite{Kretzer:2000yf} (NLO set, $Q_0=0.4$ GeV$^2$)
is also shown for comparison. 

\section{Model calculation of the Collins fragmentation function}

\begin{figure}
\includegraphics[width=10cm]{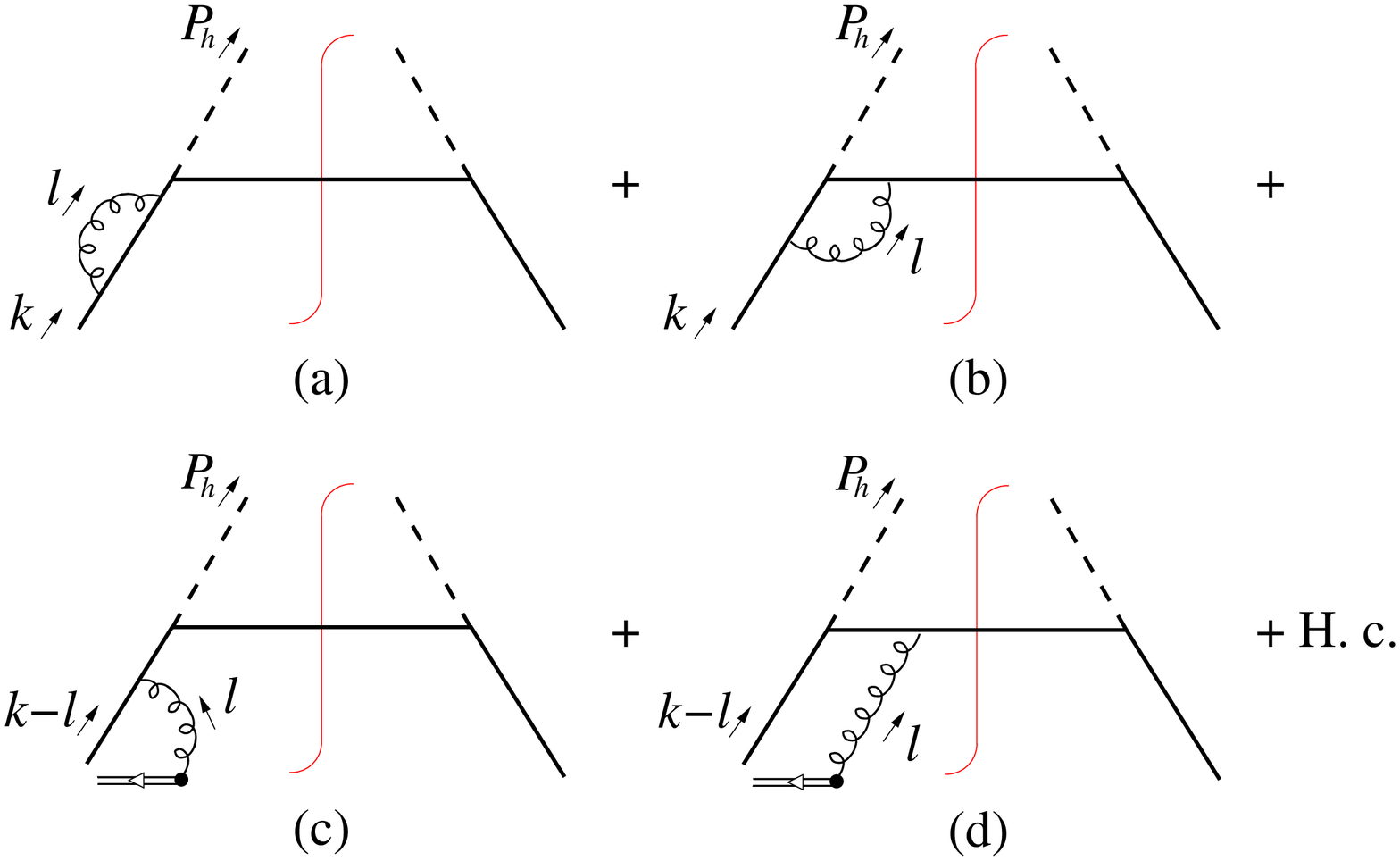}%
\caption{\label{gluonloop} Single gluon-loop corrections to the fragmentation
  of 
a quark into a pion contributing to the Collins function  in the eikonal
approximation. ``H.c.'' stands for the
hermitian conjugate diagrams which are not shown.}    
\end{figure}

We use the following definition of the Collins function~\cite{Amrath:2005gv}\footnote{The factor $1/2$ is due to a slightly different
  definition of the correlator in Eq.~\eqref{e:delta} 
with respect to Ref.~\cite{Amrath:2005gv}}
\begin{equation}
\frac{\epsilon_T^{ij} k_{Tj}}{M_h} H_1^\perp(z, k_T^2)=\frac{1}{2}\,{\rm Tr}[\Delta(z,
{k_T}) i \sigma^{i-} \gamma_5].
\end{equation}
As is well known~\cite{Amrath:2005gv}, 
using the tree-level calculation of the correlator function
is not sufficient to produce a non-vanishing Collins function, due to the lack
of imaginary parts in the scattering amplitude.
In order to obtain the
necessary imaginary part, we take into account  gluon loops. In fact, 
gluon exchange is essential to ensure color gauge invariance of the
fragmentation functions. 
Contributions come from the four diagrams in
Fig.~\ref{gluonloop}.   
Diagrams (a) and (b) 
represent the quark self-energy and vertex diagrams, respectively. 
Diagrams (c) and (d) 
can be called hard-vertex and box diagrams,
respectively. For the calculation of the diagrams with the
eikonal line, the Feynman rules to be used are
\begin{align}
   \includegraphics[width=0.7cm]{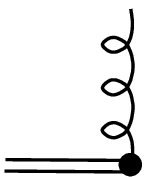}\hspace{2mm} &= \ii g \, t^{a}\, \delta^-_{\nu},
&
 \includegraphics[width=1cm]{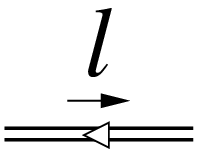} &= \frac{\ii}{-l^- \pm\ii \epsilon},
\end{align} 
where $g$ is the QCD coupling.
Note that the sign of the $\ii \epsilon$ in the eikonal propagator is
different for semi-inclusive deep inelastic scattering $(+)$ and $e^+e^-$
annihilation $(-)$, but this does not influence the computation of the Collins
function~\cite{Metz:2002iz,Collins:2004nx}.
The resulting formula for the fragmentation correlator corresponding
to Fig.~\ref{gluonloop} (d) is
\begin{equation} \begin{split} 
\Delta_{1(d)}(k,p) &= \frac{4\alpha_s}{(2 \pi)^3}\, \frac{\ii(\kslash + m)}{k^2 -
  m^2}\, g_{q \pi} \gamma_5
\,(\kslash -\Pslash_h +m_s)\,2\pi\, \delta((k-P_h)^2-m_s^2) \\
&\quad\int\frac{d^4 l}{(2 \pi)^4} \,
\frac{\ii \gamma_{\mu} t^a \,\ii(\kslash - \Pslash_h - \lslash +m_s)\, 
g_{q \pi}  \gamma_5\, 
\ii(\kslash -
  \lslash +m)\, \ii \,(-\ii g^{\mu -})\, (\ii t^a)}{((k-P_h-l)^2 -
  m_s^2+\ii \varepsilon ) ((k-l)^2 - m^2 +\ii \varepsilon) (-l^- \pm i
  \varepsilon)(l^2 - m_g^2 +\ii \varepsilon)}.
\end{split} \end{equation}
Note that one of the form factors $g_{q\pi}$ is inside the loop
integral. When using a form factor as in Eq.~\eqref{e:ff}, it would seem
reasonable to replace $k^2$ with $(k-l)^2$. However, since the form factor is
introduced to the purpose of cutting off the high-$k_T$ region, we prefer to
maintain the form factor depending only on $k^2$, so that it can be pulled out
of the integral and simplify the calculation. This choice is similar to
imposing a sharp cutoff on $k^2$ --- as done in
Ref.~\cite{Bacchetta:2002tk,Bacchetta:2003xn,Amrath:2005gv}  --- 
and not on $(k-l)^2$.  

The Collins function is given by (we take always $m_g=0$)
\begin{equation}
H_1^\perp (z, k_T^2)= -\frac{2\,\alpha_s g_{q\pi}^2}{(2 \pi)^4}\,C_F\,\frac{e^{-\frac{2k^2}{\Lambda^2}}}{z^2}\,\frac{M_h}{(1 - z)}\,\frac{1}{k^2 - m^2}
\Bigl(\tilde{H}_{1(a)}^\perp (z, k_T^2) + \tilde{H}_{1(b)}^\perp (z, k_T^2) + 
\tilde{H}_{1(d)}^\perp(z, k_T^2)\Bigr),
\end{equation}
where the subscripts in the r.h.s. refer to the  
contributions from diagrams \ref{gluonloop} (a), (b) and (d) plus their
Hermitean conjugate,
respectively. Diagram (c) gives no contribution to the Collins function.     
The separate contributions read, for the fragmentation of $u \to \pi^+$,
\begin{equation}
\tilde{H}_{1(a)}(z, k_T^2) = \frac{m}{k^2 - m^2}\,\biggl(3-\frac{m^2}{k^2}\biggr)\,I_{1g},
\end{equation}
\begin{equation}
\tilde{H}_{1(b)}(z, k_T^2) = 2\,m_s\,
I_{2g},
\end{equation}
\begin{equation}
\begin{split} 
\tilde{H}_{1(d)}(z, k_T^2) & = \frac{1}{2 z \bm{k_T}^2}\,
        \Bigl\{- I_{34g}\, (2 z m + m_s - m) +
    I_{2g} \Bigl[2 zm\Bigl(k^2-m^2+M_h^2(1-2/z)\Bigr)
\\ &\quad
        +(m_s-m)\Bigl((2z-1)k^2 - M_h^2 +m_s^2-2zm(m+m_s) \Bigr)\Bigr]\Bigr\}.
\end{split} 
\end{equation}
The loop integrals $I_{1g}$, $I_{2g}$ and
$I_{34g}$~\cite{Amrath:2005gv}\footnote{The expression $I_{34g}$ used here
  corresponds to Eq.~(24) in Ref.~\cite{Amrath:2005gv} multiplied by $k^-$.} 
are given by
\begin{align} 
I_{1g} &= \frac{\pi \sqrt{\lambda(m, m_g)}}{2 k^2},
\\
I_{2g} &= \frac{\pi}{2 \sqrt{\lambda(m_s, M_h)}}\,
\ln \biggl[\frac{k^2 + m_s^2 -M_h^2 - \sqrt{\lambda(m_s, M_h)}}
        {k^2 + m_s^2-M_h^2 + \sqrt{\lambda(m_s, M_h)}}\biggr],
\\
I_{34g} &= \pi\, \ln \biggl[ \frac{\sqrt{k^2} (1 - z)}{m_s} \biggr],
\end{align} 
where $\lambda(m_1, m_2) = (k^2 - (m_1 + m_2)^2)~(k^2 - (m_1 - m_2)^2)$. 
In the limit $m_s=m$ and setting the form factor to $1$, our result for 
$H_1^\perp$ reduces to Eq.~(15) of Ref.~\cite{Amrath:2005gv} (multiplied by
two because in that paper the results refer to $u \to \pi^0$).    
It is important to  note that the Collins function should obey the
positivity bound~\cite{Bacchetta:1999kz,Bacchetta:2002tk}
\begin{equation}
\frac{|\bm{k}_T|}{M_h} H_1^\perp (z, k_T^2) \le D_1 (z, k_T^2).
\end{equation}
Integration over $k_T^2$ gives the simplified expression
\begin{equation}
\frac{H_1^{\perp (1/2)} (z)}{D_1(z)} \le \frac{1}{2},
\label{intpositivity}
\end{equation} 
where the half moment of the Collins function is defined as
\begin{equation}
H_1^{\perp (1/2)} (z) = \pi z^2 \int_0^\infty dk_T^2 \, 
\frac{|\bm{k}_T|}{2 M_h}\, H_1^\perp  (z, k_T^2).
\end{equation}
\begin{figure}
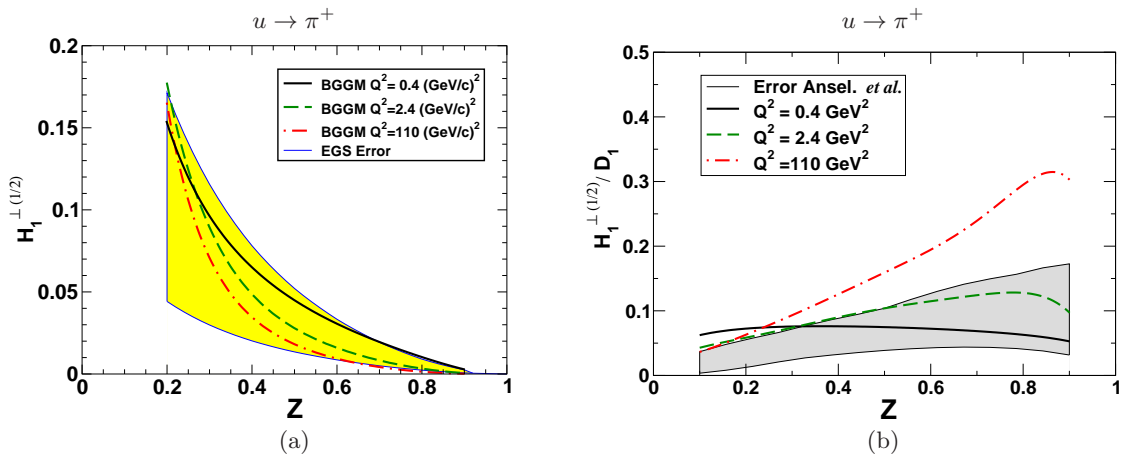

\begin{tabular}{ccc}
$u \to \pi^+$& &$u \to \pi^+$
\\
\hspace{-0.9cm}\includegraphics[height=5.1cm,clip]{half_upi+}
&
\hspace{1.5cm}
&
\hspace{-0.8cm}\includegraphics[height=5cm,clip]{ratio_upi+}
\\
(a)&&(b)
\end{tabular}
\caption{\label{collpion}Half moment of the Collins function for  
$u \to \pi^+$ in
our model. (a) $H_1^{\perp (1/2)}$ at the model scale (solid line) and at a
different scale under the assumption in
Eq.~\eqref{e:rationoevolve} (dot-dashed line), 
compared with the error band from the
extraction of Ref.~\cite{Efremov:2006qm}, (b) $H_1^{\perp (1/2)}/D_1$ at the
model scale (solid line) and at two other scales (dashed and dot-dashed lines)
under the assumption in Eq.~\eqref{e:collinsnoevolve}.
The error band from the 
extraction of Ref.~\cite{Anselmino:2007fs} is shown for comparison.}
\end{figure}
In Fig.~\ref{collpion} (a), we have plotted the half moment of the Collins
functions vs.\
z for the case $u \to \pi^+$. In the same panel, we plotted the 1-$\sigma$ 
error band of the Collins function extracted in Ref.~\cite{Efremov:2006qm} 
from BELLE data, collected at a scale $Q^2=(10.52)^2$ GeV$^2$. In order to
achieve a reasonable agreement with the phenomenology, 
we choose a value of the strong coupling
constant $\alpha_s=0.2$. Such a value is particularly small, especially when
considering that our model has been tuned to fit the function $D_1$ at a scale
$Q_0^2=0.4$ GeV$^2$, where standard NLO calculations give
$\alpha_s\approx 0.57$~\cite{Kretzer:2000yf,Gluck:1998xa}. 
In any case, the problem of the choice of $\alpha_s$ is
intimately related with the problem of the evolution of the Collins function
(see below).

In Fig.~\ref{collpion} (b), we have plotted the ratio $H_1^{\perp (1/2)}/D_1$
and compared it to the error bands of the extraction in
Ref.~\cite{Anselmino:2007fs}. Also in this case the agreement is good, with
the above mentioned choice of $\alpha_s=0.2$.

At this point, some comments are in order concerning the evolution 
of the Collins function (or of its half-moment) with the energy
scale. Such evolution is presently unknown, except for some work done in
Ref.~\cite{Henneman:2001ev}, which is however based on 
questionable assumptions. 
Some authors (e.g.\ Refs.~\cite{Efremov:2006qm,Anselmino:2007fs}) 
assume
\begin{equation}
 \frac{H_1^{\perp (1/2)}}{D_1}\biggr|_{Q_0^2} =
\frac{H_1^{\perp (1/2)}}{D_1}\biggr|_{Q^2},
\label{e:rationoevolve}
\end{equation} 
i.e., that the evolution of $H_1^{\perp (1/2)}$ is
equal to that of $D_1$. This seems unlikely, in view of 
the fact that the Collins function 
is  chiral-odd and thus  
evolves as a non-singlet.
An alternative choice could be to assume 
\begin{equation} 
H_1^{\perp (1/2)}\Bigr|_{Q_0^2} =
H_1^{\perp (1/2)}\Bigr|_{Q^2}
\label{e:collinsnoevolve}
\end{equation} 
i.e., that $H_1^{\perp (1/2)}$
does not evolve with the energy scale. This is 
an
extreme hypothesis, which cannot be true because at some point
the positivity bound \eqref{intpositivity} would be violated at large
$z$.  We demonstrate this
in Fig.~\ref{collpion} (b) where 
we show how the ratio $H_1^{\perp (1/2)}/D_1$ behaves at
at three different energy scales if only $D_1$ is evolved (we use 
the unpolarized fragmentation function of Ref.~\cite{Kretzer:2000yf} for this
purpose). Clearly, in this case the ratio grows more steeply with $z$
at higher energies, due to the decreasing of $D_1$ in the large-$z$
region. 
While the evolution of the T-odd parton distribution
and fragmentation functions remain
an outstanding issue, these results show that different
assumptions on the Collins function scale dependence have 
a significant impact and should be
considered with care.

For the fragmentation $u \to K^+$, the same analytic formulas are used but
with the other set of parameter values. 

\begin{figure}
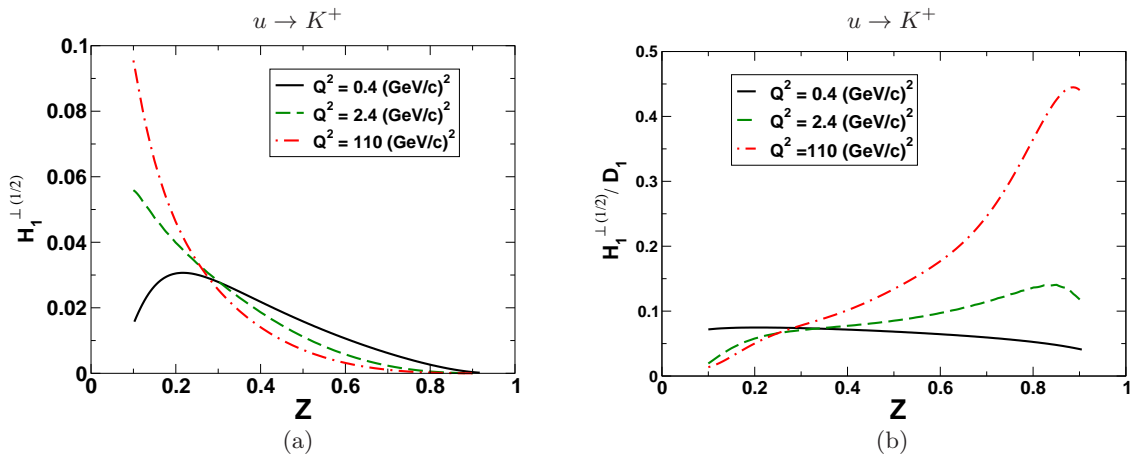

\begin{tabular}{ccc}
$u \to K^+$& &$u \to K^+$
\\
\hspace{-0.9cm}\includegraphics[height=5.1cm,clip]{half_uK+}
&
\hspace{1.5cm}
&
\hspace{-0.8cm}\includegraphics[height=5cm,clip]{ratio_uK+}
\\
(a)&&(b)
\end{tabular}
\caption{\label{colluK}Half moment of the Collins function for 
$u \to K^+$ in our model. 
(a) $H_1^{\perp (1/2)}$ at the
model scale of 0.4 GeV$^2$, (b) $H_1^{\perp (1/2)}/D_1$ at the
model scale (solid line) and at two other scales (dashed and dot-dashed
lines)  
under the assumption in Eq.~\eqref{e:collinsnoevolve}.}
\end{figure}

\begin{figure}
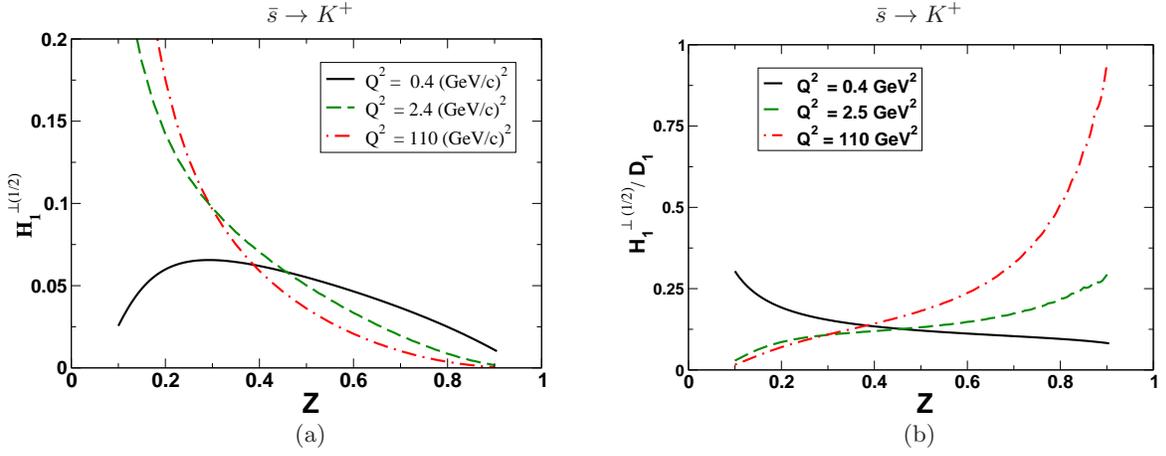

\begin{tabular}{ccc}
$\bar{s} \to K^+$& &$\bar{s} \to K^+$
\\
\hspace{-0.9cm}\includegraphics[height=5.1cm,clip]{half_sbK+}
&
\hspace{1.5cm}
&
\hspace{-0.8cm}\includegraphics[height=5cm,clip]{ratio_sbK+}
\\
(a)&&(b)
\end{tabular}
\caption{\label{collsbarK}Half moment of the Collins function for 
$\bar{s} \to K^+$ in our model. 
(a) $H_1^{\perp (1/2)}$ at the
model scale of 0.4 GeV$^2$, (b) $H_1^{\perp (1/2)}/D_1$ at the
model scale (solid line) and at two other scales (dashed and dot-dashed
lines)  
under the assumption in Eq.~\eqref{e:collinsnoevolve}.}
\end{figure}

\section{Asymmetries in $e^+e^-$ annihilation}

The BELLE collaboration has reported measurements of various asymmetries in
$e^+ + e^- \rightarrow \pi^\pm + \pi^\pm +X$ that can isolate the Collins
functions~\cite{Abe:2005zx}. In particular, the number of pions in this case
has an azimuthal dependence~\cite{Boer:1998qn}
\begin{equation} 
\begin{split}  
N_{h_1 h_2}(z_1,z_2)\propto \sum_{q}{e_q}^2 \;\Bigg(&
    D_{1 (q \to h_1)}(z_1)D_{1 (\bar q \to h_2)}(z_2)  \\
&+ \frac{\sin^2\theta}{1+\cos^2\theta}\;\cos(\phi_1+\phi_2)\;
  H_{1 (q \to h_1)}^{\perp(1/2)}(z_1)\overline H_{1 (\bar q \to
      h_2)}^{\perp(1/2)}(z_2)
  \Bigg),
\label{a12}
\end{split} 
\end{equation} 
where $\phi_{1,2}$ are the azimuthal angles of the two pions relative to their
jet axes (or thrust direction) and the 2 jet production plane. Normalizing
this distribution and extracting the azimuthal asymmetry gives a measure of
the product of moments of Collins functions. BELLE noted that there are QCD
radiative corrections that compete with the leading twist effects. To cancel
out those corrections they take the ratio of the asymmetry for unlike sign
events ($\pi^+ \pi^-$) to the asymmetry for like sign events. This super ratio
has the form~\cite{Anselmino:2007fs}
\begin{equation} 
A_{12}(z_1,z_2)=\frac{\langle \sin^2\theta \rangle}
{\langle 1+\cos^2\theta \rangle}\,(P_U-P_L )\,,
\label{A12}
\end{equation} 
where
\begin{align} 
  P_U&= \frac{\sum_q e^2_q \; \Bigl(H^{\perp (1/2)}_{1\,(q \to \pi^+)}(z_1)\,
    H^{\perp (1/2)}_{1\,(\bar q \to\pi^-)}(z_2) + H^{\perp (1/2)}_{1\,(q
      \to\pi^-)}(z_1)\, H^{\perp (1/2)}_{1\,(\bar q \to\pi^+)}(z_2)\Bigr)}
  {\sum_q e^2_q \;\Bigl(D_{1\,(q \to \pi^+)}(z_1)\, D_{1\,(\bar q
      \to\pi^-)}(z_2) + D_{1\,(q \to\pi^-)}(z_1)\, D_{1\,(\bar q
      \to\pi^+)}(z_2)\Bigr)} \,,
  \\
  P_L&= \frac{\sum_q e^2_q \; \Bigl(H^{\perp (1/2)}_{1\,(q \to \pi^+)}(z_1)\,
    H^{\perp (1/2)}_{1\,(\bar q \to\pi^+)}(z_2) + H^{\perp (1/2)}_{1\,(q
      \to\pi^-)}(z_1)\, H^{\perp (1/2)}_{1\,(\bar q \to\pi^-)}(z_2)\Bigr)}
  {\sum_q e^2_q \;\Bigl(D_{1\,(q \to \pi^+)}(z_1)\, D_{1\,(\bar q
      \to\pi^+)}(z_2) + D_{1\,(q \to\pi^-)}(z_1)\, D_{1\,(\bar q
      \to\pi^-)}(z_2)\Bigr)}.
\end{align} 
Note that Eq.~\eqref{A12} is a linear approximation for $P_L<<1$.  For
numerical studies, we take the unpolarized fragmentation functions from
Ref.~\cite{Kretzer:2000yf} (NLO set) at the scale of the BELLE measurements,
i.e., $Q^2=(10.52)^2$ GeV$^2$. We take also $\langle \sin^2\theta \rangle /
\langle 1+\cos^2\theta \rangle \approx 0.79$.

For the calculation of the asymmetry we have to make some assumptions on the
unfavored Collins fragmentation functions.
In order to have a guiding principle for our assumptions, we
consider the Sch\"afer--Teryaev sum rule~\cite{Schafer:1999kn}, which states
that
\begin{align}
  \sum_h \int_0^1 \de z H_{1 (q\to h)}^{\perp (1)}(z) &= 0 & \text{with} &&
  H_1^{\perp (1)} (z) &= \pi z^2 \int_0^\infty dk_T^2\, \frac{\bm{k}_T^2}{2
    M_h^2}\, H_1^\perp (z, k_T^2).
\end{align} 
We assume that the sum rule holds in a strong sense, i.e., for pions and kaons
separately and at the integrand level, for each value of $z$ and $k_T$.  For
pions, it follows that
\begin{equation} 
H_{1 (u \to \pi^-)}^{\perp (1/2)} = - H_{1 (u \to \pi^+)}^{\perp (1/2)}.
\end{equation} 
The other $\bar{u}$, $d$, $\bar{d}$, unfavored Collins functions are related
to the above by isospin and charge symmetries, Eq.~\eqref{unfisocharge1}. Our
strong interpretation of the Sch\"afer--Teryaev sum rule together with
Eq.~\eqref{unfisocharge2} (with $D_1$ replaced by $H_1^{\perp (1/2)}$) implies
\begin{equation}
H_{1 (s \to \pi^-)}^{\perp (1/2)} = - H_{1 (s \to \pi^+)}^{\perp (1/2)} =0.
\end{equation} 

For kaons, the same considerations lead to the following assumptions
\begin{align}
  H_{1 (u \to K^-)}^{\perp (1/2)} &= - H_{1 (u \to K^+)}^{\perp (1/2)},
  \\
  H_{1 (\bar{s} \to K^-)}^{\perp (1/2)} &= - H_{1 (\bar{s} \to K^+)}^{\perp
    (1/2)},
  \\
  H_{1 (d \to K^-)}^{\perp (1/2)} &= - H_{1 (d \to K^+)}^{\perp (1/2)} = 0.
\end{align}

\begin{figure}
\includegraphics[height=9cm]{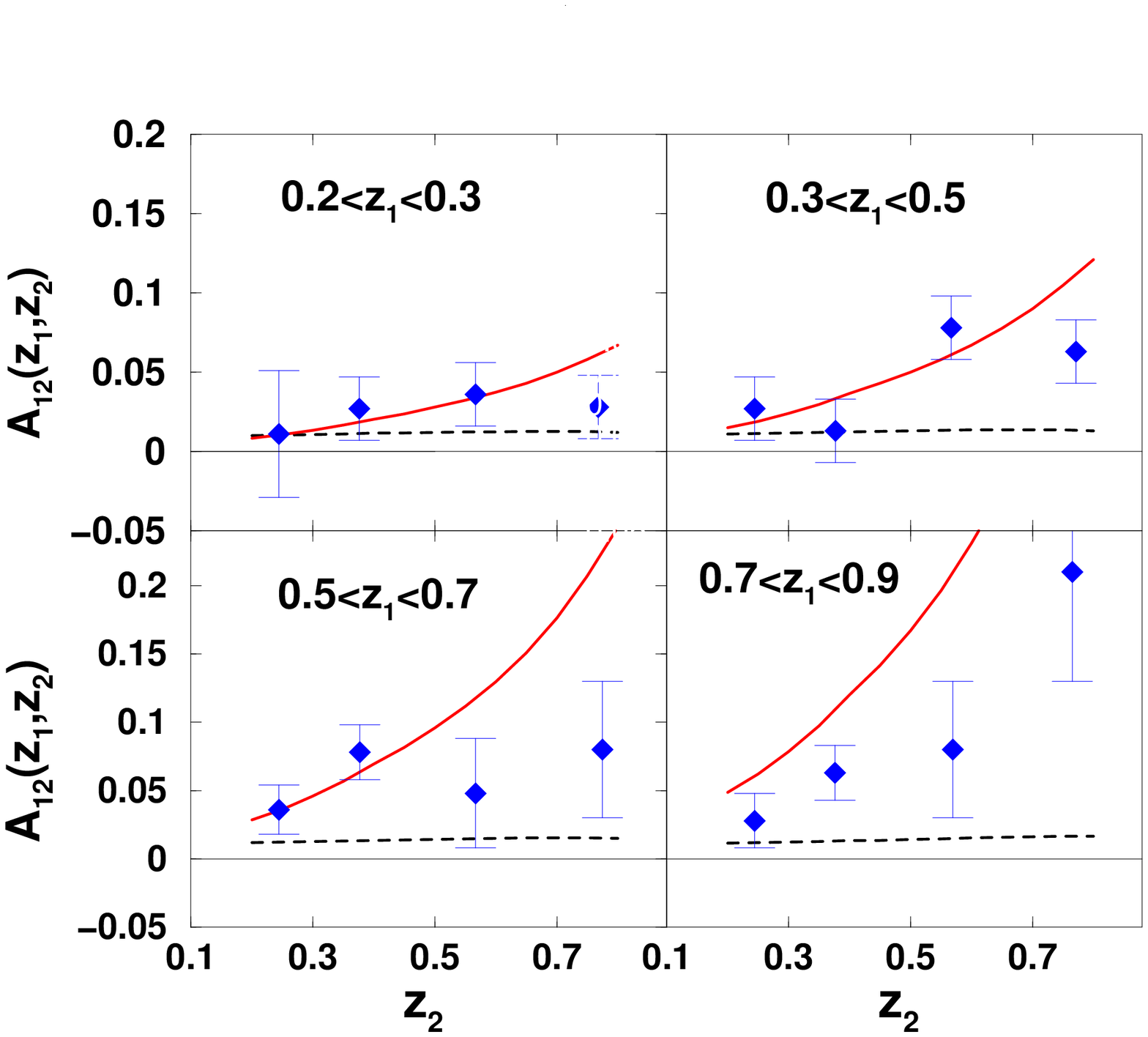}%
\caption{Azimuthal asymmetry $A_{12}(z_1,z_2)$ for the production of two pions
  as a function of $z_2$ and integrated in bins of $z_1$ at
  $Q^2=110.7~\text{GeV}^2$. Dashed lines are obtained
  assuming Eq.~\eqref{e:rationoevolve}, solid lines assuming
  Eq.~\eqref{e:collinsnoevolve}.
Note that the last $z_1$ bin in our calculation is
  narrower than in the corresponding experimental measurement.}
\label{A12_p} 
\end{figure}

In Fig.~\ref{A12_p} we show the values of the pion azimuthal asymmetry for
four different ranges of $z_1$, as a function of $z_2$. 
The dashed curves and solid curves are obtained respectively 
under the assumptions in Eq.~\eqref{e:rationoevolve} and 
Eq.~\eqref{e:collinsnoevolve}, respectively.
The upper curves 
exceed the data for the higher $z_2$ values, which either reflects
the need for corrections to the linear approximation in Eq.~\eqref{A12}, or
more likely that assuming no evolution for the Collins function may be 
too severe
an approximation.

\begin{figure}
  \includegraphics[height=9cm]{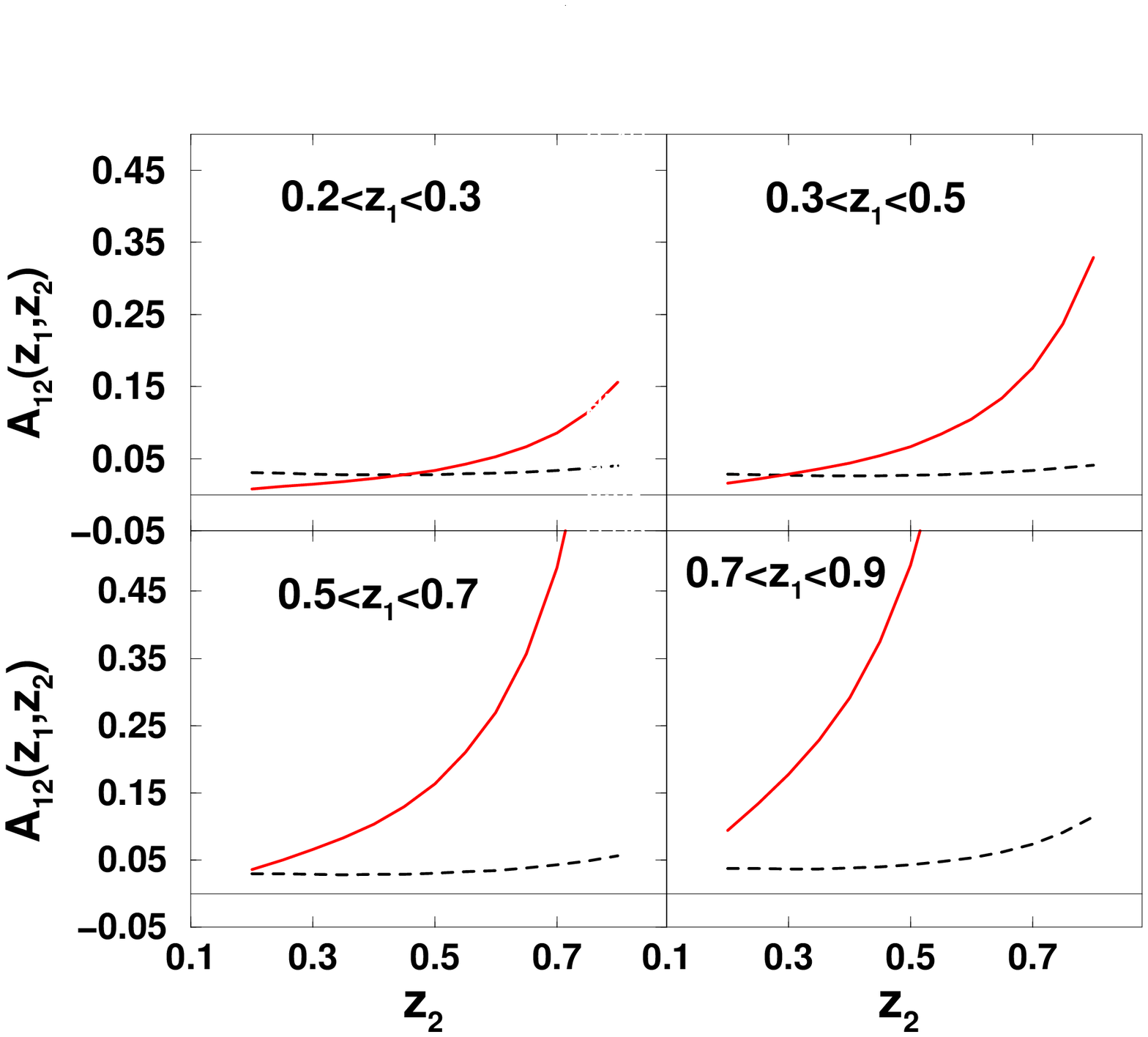}
\caption{Azimuthal asymmetry $A_{12}(z_1,z_2)$ for the production of two kaons
  as a function of $z_2$ and integrated in bins of $z_1$ at 
  $Q^2=110.7~\text{GeV}^2$. Dashed lines are obtained
  assuming Eq.~\eqref{e:rationoevolve}, solid lines assuming
  Eq.~\eqref{e:collinsnoevolve}. }
\label{A12_k} 
\end{figure}

We calculated the corresponding $K K$ asymmetry, Fig.~\ref{A12_k}, 
and obtained even
larger values, suggesting that there will be more dramatic effects in this
accessible channel.

\section{Conclusions}

In this paper, we performed a new calculation of the Collins fragmentation
function for $u \to \pi^+$, along the lines of Refs.~\cite{Bacchetta:2003xn,Amrath:2005gv} but
with some important differences: (i) we assumed the mass of the spectator is
different from the mass of the fragmenting quark, (ii) 
we introduced a form factor
at the hadron-quark vertex, (iii) 
we fitted the values of the model parameters to
reproduce the unpolarized fragmentation function $D_1$ at a scale $Q_0^2 =
0.4$ GeV$^2$. 
We compared the results of our model calculation to the available
parametrizations of the Collins
function~\cite{Efremov:2006qm,Anselmino:2007fs}
 extracted from $e^+ e^-$ annihilation
and SIDIS data and found a reasonable agreement. We stressed the importance of
critically considering different assumptions on the evolution of the Collins
function with the energy scale.

For the first time we presented an estimate of the Collins function 
for $u \to K^+$ and $\bar{s} \to K^+$. In particular, we found that the ratio 
$H_1^{\perp (1/2)}/D_1$ for $u \to K^+$ is almost identical to that for $u \to
\pi^+$, while the ratio for $\bar{s} \to K^+$ is about twice as big.

Using the results of our model, we presented estimates for pion and kaon
Collins asymmetries in $e^+ e^-$ annihilation at the BELLE experiment. 
In order to calculate the  unfavored Collins functions 
we adopted the 
``strong interpretation'' of the Sch\"afer--Teryaev
sum-rule~\cite{Schafer:1999kn}. Our results are in qualitative agreement with
the available BELLE data on the pions, 
but large uncertainties arise from making different
assumptions on the evolution of the Collins function as well as 
from determining  the 
unfavored Collins fragmentation function.
For the kaons, we predict the asymmetries to be larger than the pions.

\acknowledgments
The work of L. G and G.R.G. are 
supported in part by the U.S. Department of Energy under
contracts DE-FG02-07ER41460 and  DE-FG02-92ER40702 respectively. 
A.M. wishes to thank the hospitality of 
the DESY Theory Group, where part of the present work was carried out.  

\bibliographystyle{myrevtex}
\bibliography{mybiblio}

\end{document}